# Homogeneous quantum cascade lasers operating as Terahertz frequency combs over their entire operational regime

**Alessandra Di Gaspare,[1][*] Leonardo Viti,[1][*] Harvey E. Beere,[2] David D. Ritchie[2] and Miriam S. Vitiello[1]**

[1]*NEST, CNR - Istituto Nanoscienze and Scuola Normale Superiore, Piazza San Silvestro 12, 56127, Pisa, Italy*
[2]*Cavendish Laboratory, University of Cambridge, Cambridge CB3 0HE, UK*

\* authors contributed equally to the work

## Abstract

We report a homogeneous quantum cascade laser (QCL) emitting at Terahertz (THz) frequencies, with a total spectral emission of about 0.6 THz centered around 3.3 THz, a current density dynamic range of $J_{dr}$=1.53, and a continuous wave output power of 7 mW. The analysis of the intermode beatnote unveils that the devised laser operates as optical frequency comb (FC) synthesizer over the whole laser operational regime, with up to 36 optically active laser modes delivering ~ 200 μW of optical power per comb tooth, a power level unreached so far in any THz QCL FC. A stable and narrow single beatnote, reaching a minimum linewidth of 500 Hz, is observed over a current density range of 240 A/cm², and even across the negative differential resistance region. We further prove that the QCL frequency comb can be injection locked with moderate RF power at the intermode beatnote frequency, covering a locking range of 1.2 MHz. The demonstration of stable FC operation, in a QCL, over the full current density dynamic range, and without any external dispersion compensation mechanism, makes our proposed homogenous THz QCL an ideal tool for metrological application requiring mode-hop electrical tunability and a tight control of the frequency and phase jitter.

## Introduction

The terahertz (THz) region of the electromagnetic spectrum, loosely defined in the frequency window between 0.1 THz and 10 THz [1] attracted a renewed attention, in recent years, for targeting far-infrared applications requiring a tight control of the frequency and phase jitter of the laser modes, such as high-resolution and high-sensitivity spectroscopy [2, 3], telecommunications [4, 5] and quantum metrology [6], amongst many others.

Although relevant models for Gigahertz-Terahertz generation exploiting non-linearity in semiconductor superlattices have been recently proposed [7], electrically pumped quantum cascade lasers (QCLs) are the most efficient on-chip sources of optical frequency comb synthesizers (FCs) at THz frequencies [8-11], thanks to the wide frequency coverage [10, 12] and the inherently high optical power levels [13, 14], which allow continuous-wave (CW) emitting powers per comb tooth in the $3\mu W$ [12] - $60\mu W$ [8] range and the high spectral purity (intrinsic linewidths of ~100 Hz) [15]. Such a combination of performances makes THz QCL FCs the most suitable choice for the aforementioned applications.

In a THz QCL, the inherently high optical nonlinearity of the quantum engineered gain medium allows locking in phase the laser modes, passively [8, 16]. Specifically, the resonant third-order active material susceptibility inherently induces self-phase-locking through the four-wave-mixing (FWM) process. FWM tends to homogenize the mode spacing and consequently promotes the spontaneous proliferation of phase-locked equi-spaced optical modes.

However, to achieve a stable comb regime, the total chromatic dispersion has to be minimized over the laser operational bandwidth [8], so that the phase-mismatch from the material, the laser waveguide and the active region gain approaches zero in the QCL cavity. This has been achieved either by integrating a dispersion compensator in a homogeneous QCL active region [8] providing a negative GVD to cancel the positive cavity dispersion, or by gain medium engineering of heterogeneous multi-stacked active regions (ARs) [10, 17, 18], or an individual homogeneous AR [19], in both cases designed to have a relatively flat gain top, within which the intrinsic dispersion is small enough to preserve comb formation. Heterogeneous designs are ideal for achieving a broad spectral coverage, but they also come with some inherent disadvantages, such as the design-related difficulties associated with a proper matching of the threshold currents between the different AR modules, which can prevent the simultaneous mode proliferation over the full operation range of the QCL. On the other hand, homogenous ARs are usually engineered with a narrower gain profile, but are in principle easier to be injection-locked in the laser operation regime in which dispersion is not compensated.

Although a broad spectral coverage (0.7 THz – 1.1 THz) [8, 10, 12, 19] have been demonstrated using both a dispersion compensator [8] or through heterogeneous [10, 12] or homogeneous [8, 19] gain medium engineering, both approaches are ineffective in handling the bias-dependent cavity dispersion, meaning that spontaneous comb operation has been demonstrated to occur only over a limited driving current dynamic range, usually < 20-25% of the QCL operational regime [8-10, 12, 19]. To date, external approaches such as coupled $dc$-biased cavities [20] or Gires–Tournois interferometers [21], proved to be the only way to increase the operational dynamic range of THz FC synthesizers, although with poor spectral coverage [20], optical power outputs [20] or with only a moderate increase of the current dynamic range [21].

In this work, we quantum engineer and devise a broadband homogeneous QCL FC covering a bandwidth of 0.6 THz (3.05 THz-3.65 THz), with a dynamic range $J_{dr} = J_{max}/J_{th}$ =1.53 ( $J_{th}$ is the threshold current density and $J_{max}$ the maximum current density value) and a maximum CW output power of 7 mW. Remarkably, the collected electrical intermode beatnote map reveals spontaneous frequency comb operation over the entire operational range of the laser, classifying the proposed device as a unique frequency-tunable metrological tool across the far-infrared.

**Results and discussion**

The active region design of the homogeneous gain medium is a slightly-modified version [22-23] of the four quantum well bound-to-continuum structure described in ref. [24]. The highly diagonal laser transition ensures gain recovery times much larger (> 35 ps) [25] than those usually achieved in THz QCLs (5-10 ps), together with shorter upper state lifetimes, above threshold. This is an ideal condition for amplitude modulations up to tens of GHz, ideal for FC operation [19]. Indeed, a GHz modulation envelope possibly synchronized to the THz field in the time domain could also help stabilizing the laser in a fashion reminiscent of active mode-locking, supporting the formation of an amplitude modulated comb. The GaAs/AlGaAs hetoerostructure, used in this work is grown by molecular beam epitaxy (MBE) on a semi-insualting GaA subtrate. The final gain medium includes a sequence of 160 periods of a 4-wells quantum cascade design centered at 3.3 THz. The 10-μm thick final structure consists of a 250 nm undoped GaAs buffer layer, an undoped 250 nm Al$_{0.5}$Ga$_{0.5}$As etch-stop layer, a 700 nm Si-doped ($2\times10^{18}$ cm$^{-3}$) GaAs layer, the AR (doped at $3\times10^{16}$ cm$^{-3}$), and a 80 nm heavily Si-doped ($5\times10^{18}$ cm$^{-3}$) GaAs top contact layer.

Device fabrication is based on a standard metal-metal processing technique that relies on Au-Au thermo-compression wafer bonding of the MBE sample onto a highly doped GaAs substrate. A Cr/Au (10nm/150nm) top contact is lithographically patterned on the top laser surface, leaving uncoated the two sides of the ridge along two lateral stripes, whose width ranges between 3 μm and 5 μm, depending on the ridge width [10]. Laser bars are then realized via deep RIE etching with vertical sidewalls to allow uniform current injection and ridge widths in the 50-90 μm range. A 5-nm thick Ni layer is then deposited on the uncoated top lateral stripes, to define lossy side absorbers, needed to suppress undesired high order lateral modes. Laser bars 50-70 μm wide and 2.75 mm long were then In-soldered to a copper plate, wire-bonded and mounted on the cold head of a helium flow cryostat.

The voltage-current density (V-J) and the light-current density (L-J) characteristics, measured while driving the QCL in CW as a function of the heat sink temperature ($T_H$), show laser action up to a maximum heat sink temperature $T_H = 88$ K. Figure 1A shows the V-J-L collected in the 15 K-75 K interval. At $T_H = 27$ K the CW threshold current density is $J_{th} = (428$ Acm$^{-2})$, the maximum current density $J_{max} = 653$ Acm$^{-2}$, the operational dynamic range is $J_{dr} = J_{max}/J_{th} = 1.53$ and the maximum peak optical power is 7 mW.

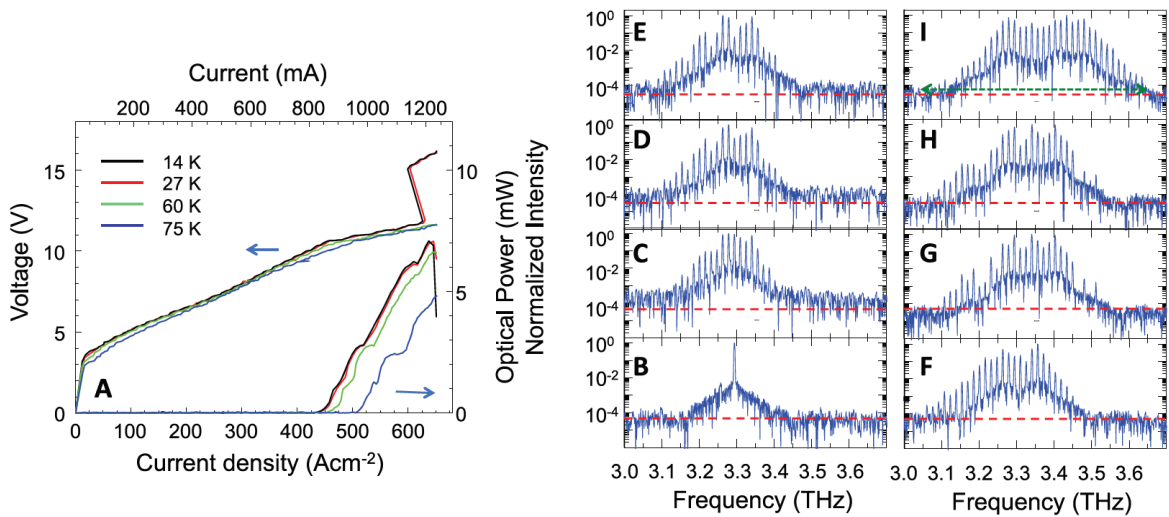

**Figure 1. (A)** Voltage- current density (V-J) and light-current density (L-J) characteristics, measured in continuous wave, as a function of the heat sink temperature in the 15-75 K range. The light blue dots indicate three prototypical operational transport regimes. **(B-I)** Fourier transform infrared spectra measured at $T_H = 27$K, in rapid scan mode, under-vacuum with a 0.075 cm$^{-1}$ spectral resolution, while driving the QCL in continuous wave at (B) 445 Acm$^{-2}$, (C) 465 Acm$^{-2}$, (D) 470 Acm$^{-2}$, (E) 480 Acm$^{-2}$, (F) 512 Acm$^{-2}$, (G) 550 Acm$^{-2}$, (H) 583 Acm$^{-2}$, and (I) 645 Acm$^{-2}$. The red dashed lines indicate roughly the noise floor of the measurements. The green arrow marks the laser bandwidth.

CW Fourier Transform Infrared (FTIR) spectra, acquired under vacuum (Bruker vertex 80) with a 0.075 cm$^{-1}$ spectral resolution (Figure 1B-I), while progressively increasing the driving current, show that the laser is initially single-mode (∼3.3 THz) at threshold (J = 445 Acm$^{-2}$, Fig. 1B) and then turns multimode with a progressively richer (Figs. 1C-I) sequence of equidistant optical modes, spaced by the cavity round trip frequency. The overall spectral coverage reaches 600 GHz (Fig. 1I), and differently from the very first demonstration [8], does not show any spectral dip; at a current density of 645 Acm$^{-2}$ (Fig. 1I), corresponding to 7 mW peak optical power, 36 optically active laser modes are retrieved. Under this condition the CW optical power per comb tooth is ∼ 200 μW, significantly larger than that of any previous THz QCL frequency comb, demonstrated so far [8-12, 19].

To investigate the coherence properties of the devised THz QCL we then trace the intermode beatnote radio frequency (RF) as a function of the driving current (Fig. 2A). The QCL is driven in CW by a low-noise power supply (Wavelength Electronics QCL 2000) and the radio frequency (RF) signal is recorded using a bias-tee (Tektronix AM60434) connecting the QCL and a RF spectrum analyzer (Rohde and Schwarz FSW43). Remarkably, the intermode beatnote map shows a single narrow beatnote over the entire dynamic range of the QCL, with the appearance of an individual beatnote for a current density just above threshold, persisting continuously upon entering in the negative differential resistance (NDR) region.

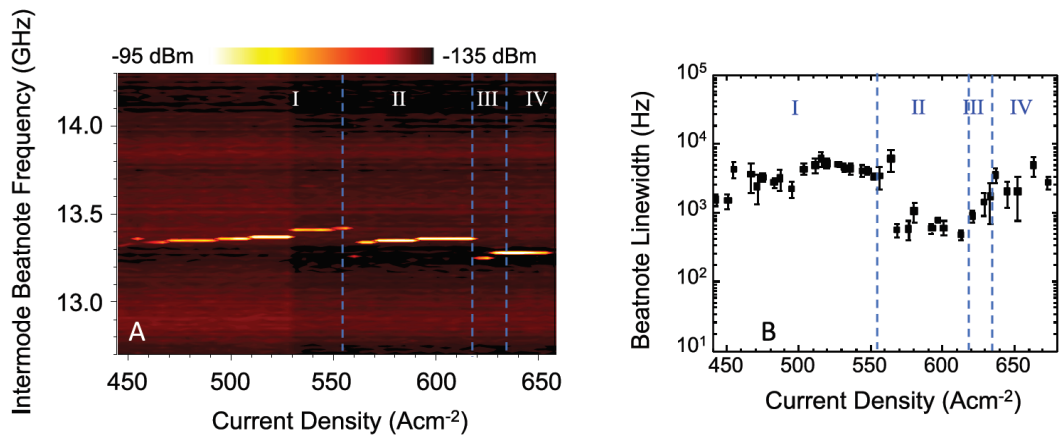

**Figure 2. (A)** Intermode beatnote map as a function of the driving current density measured at 27 K in a 2.75-mm-long, 70-μm-wide laser bar, operating in continuous wave (CW). The beatnote signal is extracted from the laser bias line using a bias-tee, and is recorded with an RF spectrum analyzer (Rohde & Schwarz FSW; resolution bandwidth (RBW): 500 Hz, video bandwidth (VBW): 500 Hz, sweep time (SWT): 500 ms, RMS acquisition mode). The vertical dashed lines identify the beginning of four relevant transport regimes: (I) onset for conduction; II (band alignment); III (peak optical power); IV (NDR)

The analysis of the beatnote linewidths (LWs) (Figure 2B) shows that the LW is initially varying in the range of ∼ 2-5 kHz (Figure 3A) for $J$ up to 558 Acm$^{-2}$ (region I in the beatnote map in Figures 2A,B), then it becomes narrower (∼500 Hz, Figure 3B) for current densities up to 616 Acm$^{-2}$, (region II). This is followed by a limited current density region, around the peak optical power ($J$ = 632 Acm$^{-2}$) (region III), in which the LW increases progressively from 500 Hz up to 2 kHz. Interestingly, when the QCL is then driven beyond the roll-off current (region IV), the beatnote is still narrow (2-7 kHz, Figure 3C), but weaker (∼15 dBm), due to the electric field instabilities, associated with the bias fluctuations, occurring in this regime. Surprisingly, the laser does not show any high phase noise regime [8-12, 19, 21].

The analysis of the electrical frequency tuning of the intermode beatnote reflects a different dynamics along the distinctive operational bias regimes. Indeed, by varying the driving current at a fixed operating temperature $T_H$ = 27 K, the beatnote initially blue shifts across region I, spanning a 70 MHz range with a tuning coefficient of +0.47 MHz/mA, then the beatnote jumps at a lower value at the onset of region II and the tuning coefficient then decreases to +0.2 MHz/mA. The beatnote then jumps again at a lower value (13.315 GHz) at the onset of region III and remains almost constant over the entire NDR regime. The overall tuning behavior clearly reflects the lack of Joule heating related effects that would led to a negative tuning coefficient with a monotonic decrease of the beatnote frequency with respect to the driving current. The observed trend is instead ascribed to the intracavity dynamics of the AR, specifically to the chromatic dispersion affecting the frequency, and thus the spacing of the cavity modes, owing to the variation in the effective refractive index of the gain medium.

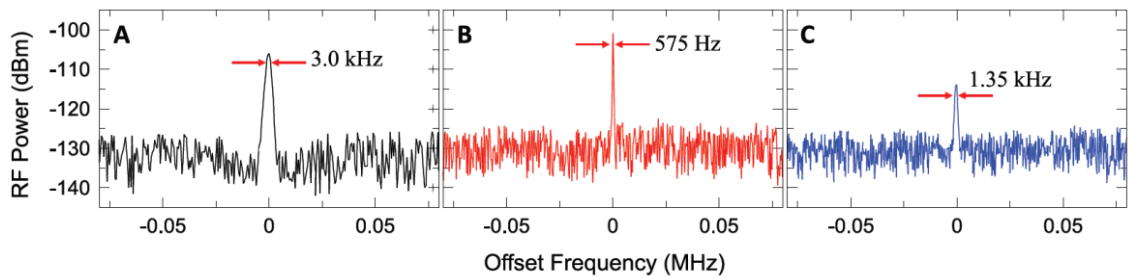

**Figure 3. (A-C)** Intermode beat note linewidth measured at 27 K, while driving the QCL in continuous wave at (A) 480 Acm$^{-2}$, (B) 570 Acm$^{-2}$, (C) 645 Acm$^{-2}$ with an RF spectrum analyzer (RBW: 500 Hz, VBW: 500 Hz, SWT: 200 ms).

To get a deeper insight on the intracavity dynamics, we perform numerical simulations of the group delay dispersion (GDD). The dispersion profile is retrieved including the contributions from the material, the waveguide and the gain of the QCL [9, 21]. The first two terms are computed considering a Drude-Lorentz model for the frequency dependent refractive index of the material; we then compute the QCL gain from the experimental emission spectra and evaluate the refractive index deviation as a consequence of the gain by applying the Kramers-Kronig equations. Finally, the dispersion provided by the whole structure is computed from the second derivative of the phase. Figure 4 shows the individual GDD contribution calculated at J = 645 Acm$^{-2}$, i.e. in the high voltage regime, in which the laser spectrum is particularly rich of optical modes.

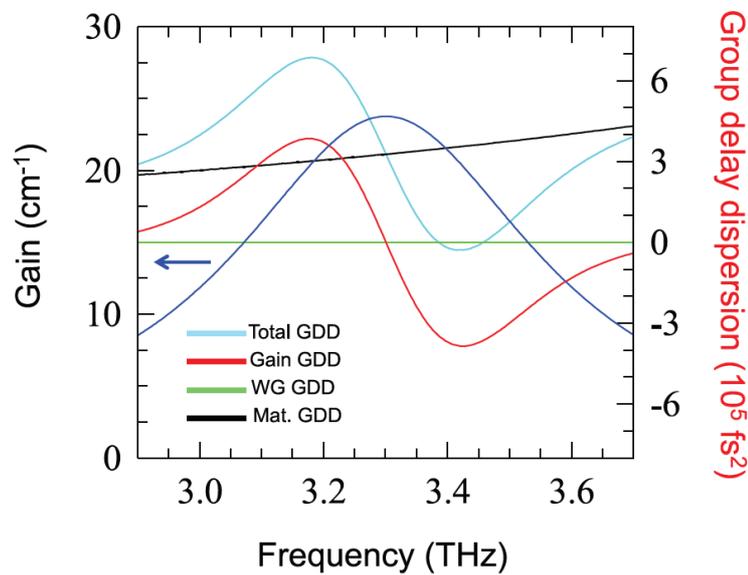

**Figure 4.** Simulations of the group delay dispersion including contributions from the material (GaAs), waveguide and gain, performed at J = 645 Acm$^{-2}$. The estimated gain curve (blue) is plotted on the left vertical axis of the graph and marked with an arrow.

In the spectral region where the QCL shows laser action, the waveguide shows a low dispersive refractive index, therefore the waveguide-related contribution to the dispersion is negligible (< 10$^5$ fs$^2$) with respect to the other terms. The gain profile (Fig. 4), centered at 3.3 THz, is extracted from the emission spectra of figure 1I, and then normalized considering the cavity length, the light intensity ratio between the center of the bandwidth and the peripheral lasing modes, and adding up the total losses from the waveguide (~ 7.3 cm$^{-1}$), the material (~ 3.0 cm$^{-1}$) [9] and the QCL facets (~ 1.6 cm$^{-1}$). The total GDD shows a visible oscillating behavior at the center of the laser gain bandwidth; in the operational range of the QCL its value ranges between $3 \times 10^5$ fs$^2$ to $2.5 \times 10^5$ fs$^2$, reaching a maximum of $6.5 \times 10^5$ fs$^2$ at 3.2 THz and approaches zero over a range of about 200 GHz. On average, even in such a high-voltage regime, the GDD is equivalent to the values reported in heterogeneous frequency combs, just above threshold, i.e. in the regime in which the QCL clearly behaves as a comb [9, 26].

We then investigate the frequency locking characteristics of the laser structure at the onset of the NDR range. Injection locking maps are acquired by using a radio frequency synthesizer (Rohde and Schwarz SMA100B) connected to the laser driver line, for the simultaneous supplying of both the CW *dc* bias and the RF injection signal over a bias. The RF and *dc* bias are fed to the QCL through 18 GHz-cutoff SMA cables and connectors up to the cryostat input ports. The simultaneous monitoring of the injection signal and of the beatnote was performed positioning a RF antenna in close proximity (~3 cm) of the QCL and recording the antenna signal through the spectrum analyzer.

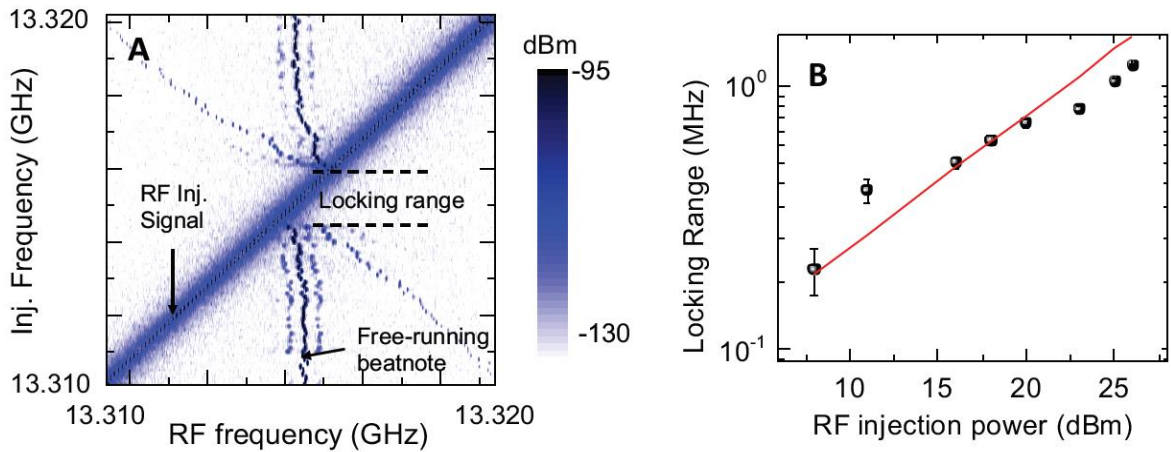

**Figure 5. (A)** Injection locking map for an injected RF power of +25 dBm. The map is recorded by sweeping the RF injection signal (diagonal line across the panel) towards the free running beatnote and then locking it. The QCL is 2.7 mm long and 70 μm wide. **(B)** Locking range extracted from the QCL injection locking at different RF injection power (black squares), following a 0.5-slope dependence in log-log scale (red line).

The injection-locking map, acquired with a RF injection power of 25 dBm and while the QCL is driven at the onset of the NDR region ($J = 635$ Acm$^{-2}$), is reported in Figure 5A. Under these conditions, the emission spectra of the QCL, under injection locking, comprises more than 36 optically active modes. The collected map discloses the typical injection locking behavior, with the initial pulling and then locking of the beatnote to the injection signal, with the simultaneous appearance of multiple sidebands.

In Figure 5B we report the locking range as a function of the injection power. The plot shows a good agreement with the 0.5-slope dependence in log-log scale foreseen in the Adler's equation [27] particularly in the lowest power region. We observed a maximum locking range of 1.2 MHz with a RF power of 25 dBm. The locking range values are slightly lower with that recently reported for homogeneus QCL frequency combs [19, 28], around the peak optical power. Conversely, the injection locking power is significantly higher than the typical values reported to date [19, 29, 28], while the actual power injected to the QCL is much lower. This can be understood by considering the total RF losses from cables, connectors, Au pads and bonding wires; by directly measuring the RF power transmitted to the cold finger we estimate a total RF attenuation of ~ 35 dB . This latter value is extracted from a direct measurement of the RF power transmitted to the cold finger.

**Conclusions**

In conclusion, we demonstrate a THz QCL based on a homogenous active region design with an emission bandwidth of 600 GHz, a CW optical output power of 7 mW, operating as frequency comb synthesizer over the entire laser operation range, including the NDR regime. The electrical intermode beatnote map unveils stable comb operation over the entire dynamic range $J_{max}/J_{th}$ = 1.45, with a single narrow beatnote reaching minimum linewidth = 500 Hz under free running operation. The laser shows more than 36 optically active, equally spaced modes delivering 200 μW of optical power per comb tooth, the largest value reported to date in any THz QCL FC. We furthermore prove the RF injection locking capability of the devised laser bars, in a regime conventionally characterized by electric field instabilities. The achieved results provide a concrete route for the development of miniaturized homogeneous chip-scale FC spectroscopic setups for addressing metrological-grade applications in the far-infrared, addressing absorption line-strengths comparable or even stronger than fundamental, mid-infrared vibration transitions, but with much narrower Doppler-limited line-widths, ruled by inverse linear relationship with the wavelength [3].

**Acknowledgements**

The authors acknowledge financial support from the ERC Project 681379 (SPRINT) and the EU union project MIR-BOSE (737017).  The authors acknowledge useful discussions with Valentino Pistore.